\documentclass[10pt, conference, letterpaper]{ IEEEtran}
\IEEEoverridecommandlockouts
\usepackage{cite}
\usepackage{amsmath,amssymb,amsfonts}
\usepackage{graphicx}
\usepackage{textcomp}
\usepackage{xcolor}
\usepackage{verbatim} 
\usepackage{mathtools}
\usepackage{filecontents}
\usepackage{tcolorbox}
\usepackage{algpseudocode}
\usepackage{booktabs}
\usepackage{times,cite,amsmath,amssymb,amsfonts,epsfig,subfigure, xcolor, soul}
\usepackage{url}
\usepackage{algorithm}
\usepackage{float}

\def\BibTeX{{\rm B\kern-.05em{\sc i\kern-.025em b}\kern-.08em
    T\kern-.1667em\lower.7ex\hbox{E}\kern-.125emX}}
\begin{document}

\title{CTMap: LLM-Enabled Connectivity-Aware Path Planning in Millimeter-Wave Digital Twin Networks\\
}

\author{
\IEEEauthorblockN{
Md~Salik~Parwez\IEEEauthorrefmark{1},
Sai~Teja~Srivillibhutturu\IEEEauthorrefmark{1},
Sai~Venkat~Reddy~Kopparthi\IEEEauthorrefmark{1},\\
Asfiya~Misba\IEEEauthorrefmark{1},
Debashri~Roy\IEEEauthorrefmark{1},
Habeeb~Olufowobi\IEEEauthorrefmark{1},
Charles~J.~Kim\IEEEauthorrefmark{2}
}
\IEEEauthorblockA{\IEEEauthorrefmark{1}Department of Computer Science and Engineering, The University of Texas at Arlington, Arlington, TX, USA\\
Email: mdsalik.parwez@uta.edu;\; \{sxs5722, sxk3962, axm8239\}@mavs.uta.edu;\\
debashri.roy@uta.edu;\; habeeb.olufowobi@uta.edu}
\IEEEauthorblockA{\IEEEauthorrefmark{2}Department of Electrical Engineering and Computer Science, Howard University, Washington, DC, USA\\
Email: ckim@howard.edu}
}

\maketitle

\begin{abstract}
In this paper, we present \textit{CTMap}, a large language model (LLM)-empowered digital twin framework for connectivity-aware route navigation in millimeter-wave (mmWave) wireless networks. Conventional navigation tools optimize only distance, time, or cost, overlooking network connectivity degradation caused by signal blockage in dense urban environments. The proposed framework constructs a digital twin of the physical mmWave network using OpenStreetMap, Blender, and NVIDIA Sionna’s ray-tracing engine to simulate realistic received signal strength (RSS) maps. A modified Dijkstra’s algorithm then generates optimal routes that maximize cumulative RSS, forming the training data for instruction-tuned GPT-4-based reasoning. This integration enables semantic route queries such as “find the strongest-signal path” and returns connectivity-optimized paths interpretable by users and adaptable to real-time environmental updates. Experimental results demonstrate that CTMap achieves up to tenfold improvement in cumulative signal strength compared to shortest-distance baselines, while maintaining high path validity. The synergy of digital twin simulation and LLM reasoning establishes a scalable foundation for intelligent, interpretable, and connectivity-driven navigation, advancing the design of AI-empowered 6G mobility systems.
\end{abstract}

\begin{IEEEkeywords}
Digital twin, large language models (LLMs), millimeter-wave (mmWave) networks, connectivity-aware navigation, received signal strength (RSS), 6G, smart cities, extended reality (XR)
\end{IEEEkeywords}

\section{Introduction}

\subsection{Background and Motivation}
Enabling seamless mobility and reliable wireless connectivity in dense urban environments remains a major challenge for future smart cities. Emerging immersive applications such as augmented, virtual, and extended reality (AR/VR/XR) demand ultra-high data rates and low latency to sustain interactive user experiences. For nomadic users, these applications offer real-time, location-aware services but impose stringent requirements on network continuity and spatial intelligence.

Millimeter-wave (mmWave) networks provide multi-gigabit bandwidth and low latency but are highly susceptible to blockage and path loss from buildings and vehicles, leading to signal degradation and frequent handovers. Consequently, traditional navigation tools such as Google Maps, Apple Maps, Waze, and HERE WeGo etc.~\cite{googlemap,applemap,waze,herewego} that prioritize shortest-distance routes often direct users through regions of poor signal coverage, causing disruptions in latency-sensitive XR or autonomous driving applications.

As illustrated in Module (1) of Fig.~\ref{flow-diagram}, consider a user equipped with an XR headset or a connected vehicle operating in a dense mmWave environment. A conventional shortest-distance route may traverse blockage-prone zones where the mmWave link intermittently drops. In contrast, a connectivity-aware navigation system would recommend a path that maximizes received signal strength (RSS) while maintaining feasible travel distance.

In this envisioned 6G scenario, the user issues a semantic query such as “Find the best-connected route from point A to B.” The physical mmWave network updates its digital twin (DT), which continuously maintains a real-time RSS map. The proposed CTMap framework queries this digital replica to infer the path that optimizes connectivity while ensuring spatial feasibility. This interaction exemplifies the fusion of digital-twin intelligence and LLM-based reasoning envisioned for 6G-enabled smart mobility.

\subsection{Related Work}
\paragraph{Digital-twin modeling for wireless networks}

Digital twins are becoming integral to 6G system design by enabling real-time emulation, control, and optimization of wireless environments.
Masaracchia et al.~\cite{masaracchia2022dt6g} and Bariah et al.~\cite{bariah2023dtcomms} introduced DT-driven communication frameworks, while Shen et al.~\cite{shen2022holistic} emphasized pervasive network intelligence.
Tao et al.~\cite{tao2024dtgenai} identified generative AI as a key driver for DT evolution.
These studies, however, focus primarily on network-level orchestration and resource management, leaving open the question of how DTs can directly support mobility-aware, signal-driven routing.

\paragraph{LLMs for network reasoning and automation}
Recent work has demonstrated that large language models (LLMs) can interpret, plan, and automate network control tasks.
Xu et al.~\cite{xu2024llm6g}, Huang et al.~\cite{huang2025llmnet}, and Long et al.~\cite{long2025llmintel} explored their roles in perception, reasoning, and alignment within 6G ecosystems.
Frameworks such as NetOrchLLM~\cite{abdallah2024netorchllm} and LLM-assisted deployment~\cite{sevim2024llmdeploy} leverage natural-language prompts for network management but lack spatial grounding and integration with propagation environments.
CTMap uniquely connects instruction-tuned LLM reasoning with DT-based signal modeling to achieve wireless-aware route generation.

\paragraph{RSS-driven localization and navigation}
RSS-based methods have long been used for indoor localization and navigation via Wi-Fi, Bluetooth, or UAV-assisted systems~\cite{kasantikul2015wifi,park2023drss,au2013indoor,chowdhury2019uav}.
However, most prior studies remain confined to small-scale or static environments and do not leverage real-time DT synchronization or semantic reasoning for outdoor mmWave scenarios.

Collectively, these gaps motivate a new paradigm that fuses DT-based signal intelligence, RSS-driven optimization, and LLM-grounded spatial reasoning, which is the foundation of CTMap.

\subsection{Problem Gap and Contributions}
Although prior research has advanced digital-twin modeling, signal-aware routing, and AI-driven network intelligence, these domains have evolved largely in isolation. Current DT frameworks focus on network emulation and control rather than user-centric connectivity guidance~\cite{masaracchia2022dt6g,bariah2023dtcomms,shen2022holistic}.
Meanwhile, recent LLM-based frameworks~\cite{abdallah2024netorchllm,huang2025llmnet,quan2025radiomap} demonstrate high-level reasoning but remain unanchored to real-world signal dynamics.
RSS-based navigation methods~\cite{kasantikul2015wifi,park2023drss,au2013indoor,chowdhury2019uav} are limited to static or indoor use cases, without leveraging DT synchronization.

To address this gap, this paper introduces CTMap, a unified digital-twin and LLM-driven framework for connectivity-aware route optimization in mmWave networks. The major contributions are:

\begin{itemize}

\item \textbf{Digital-Twin-Driven Signal Modeling:}
A realistic mmWave DT built with Blender and NVIDIA Sionna provides continuous propagation mapping and synchronization with the physical environment.
\item \textbf{Signal-Aware Path Optimization:}
A modified Dijkstra’s algorithm computes routes maximizing cumulative RSS, achieving deterministic yet connectivity-optimized navigation.
\item \textbf{Instruction-Tuned LLM Reasoning:}
A fine-tuned GPT-4 model interprets natural-language queries and infers signal-aware routes from DT data, enabling explainable and adaptive mobility guidance.
\end{itemize}
Together, these components form a unified framework for LLM-guided, DT-synchronized, connectivity-aware routing, bridging wireless network intelligence and spatial navigation to enable next-generation XR, vehicular, and smart-city applications.
\begin{figure}[t]
\centering
\includegraphics[width=0.9\linewidth]{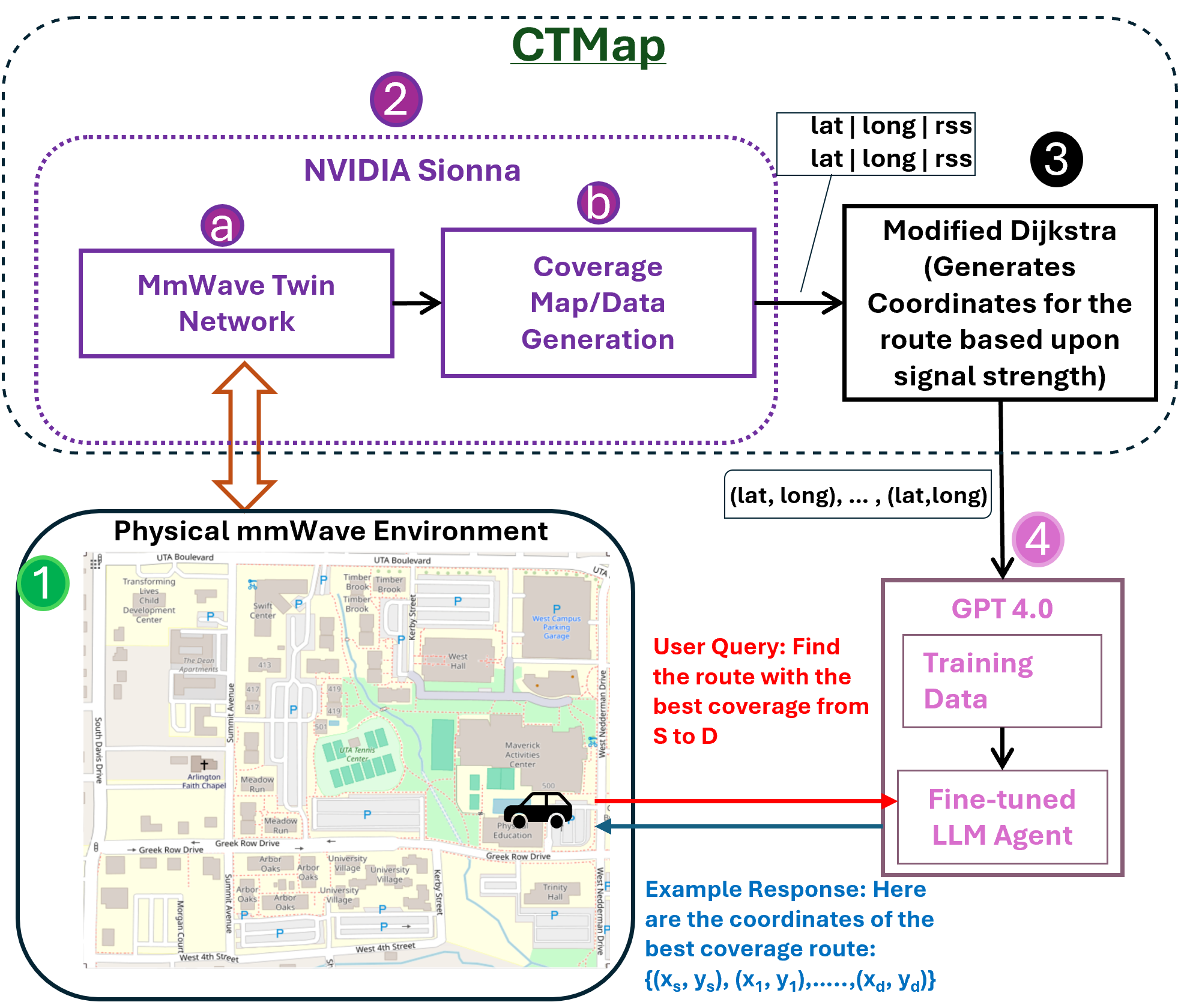}
\caption{Overall CTMap workflow integrating DT signal modeling and LLM reasoning.}
\label{flow-diagram}
\end{figure}

\section{CTMap Architecture}

This section presents the overall architecture and workflow of the proposed \textit{CTMap} framework.
%highlighting how connectivity-aware routing is realized through digital twin (DT) modeling, signal-aware optimization, and LLM-based reasoning.

\subsection{Framework Overview}
The proposed framework operates within a cloud-hosted DT environment that mirrors the physical mmWave urban network. 
Figure~\ref{flow-diagram} illustrates the four functional modules: 
(1)~Physical Environment, 
(2)~Digital Twin Construction and Signal Modeling, 
(3)~Connectivity-Aware Path Generation, and 
(4)~LLM-based Semantic Reasoning. 
Modules~(2)-(4) reside in the DT cloud, which performs continuous updates and optimization, while Module~(1) corresponds to the real mmWave deployment where users request routes semantically through CTMap.

The physical environment corresponds to a $0.7\times0.6\,\mathrm{km^2}$ urban campus extracted from OpenStreetMap (OSM)\cite{openstreetmap} and digitized using OSMnx. 
The DT replica is modeled in Blender \cite{blender} to capture 3D geometry and imported into NVIDIA Sionna\cite{sionna}, a ray-tracing engine for realistic mmWave propagation simulation. 
Each spatial location $v$ is annotated with a received signal strength value $S(v)$, producing a coverage map that serves as the input to subsequent optimization.

\begin{figure}[t]
\centering
\includegraphics[width=0.44\textwidth]{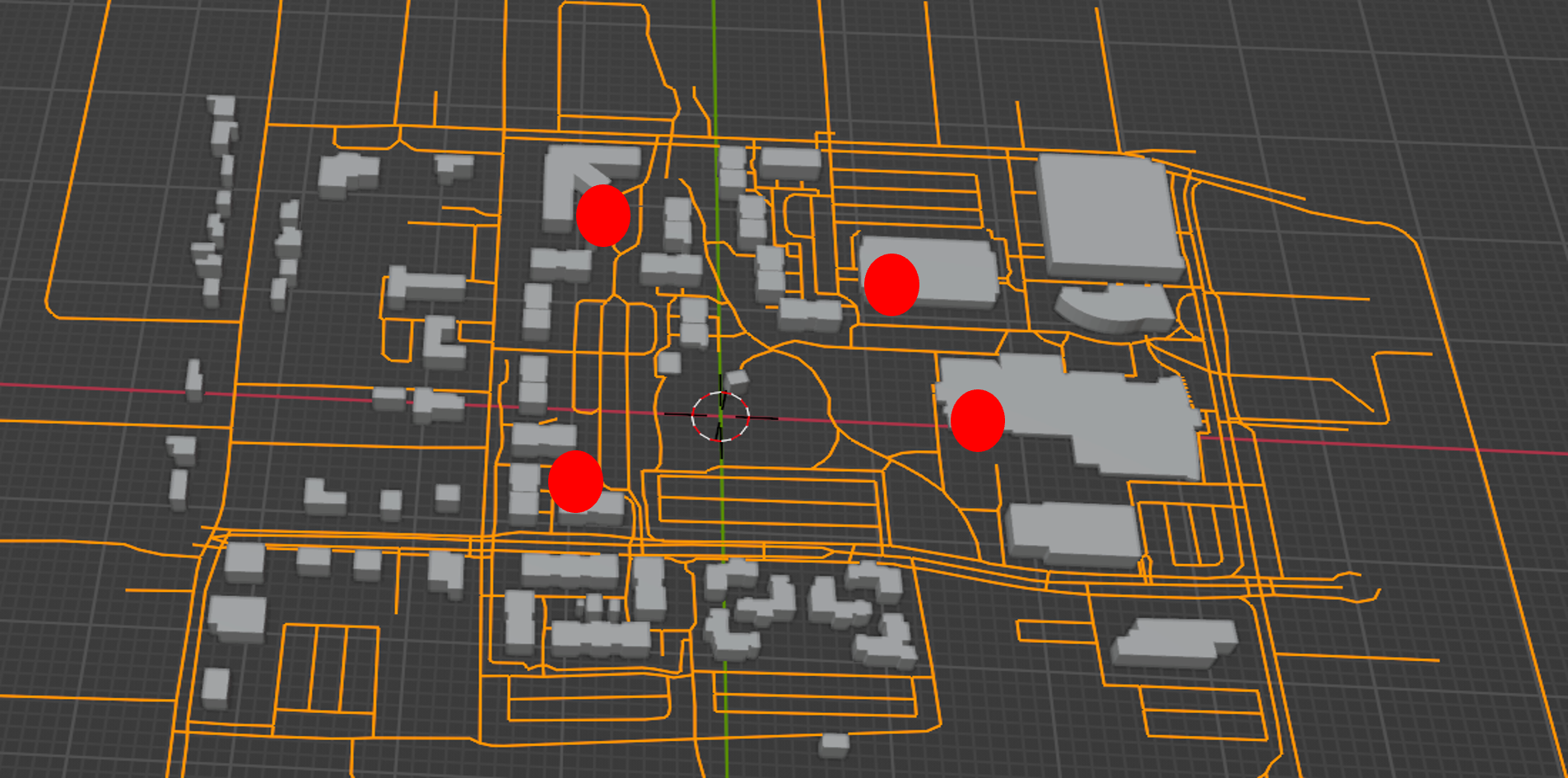}% <-- replace with your filename
\caption{Considered $0.7{\times}0.6$~km$^2$ urban campus extracted from OpenStreetMap used to build the digital-twin environment.}
\label{fig:study_area}
\end{figure}

To better illustrate the spatial domain used for simulation, Fig.~\ref{fig:study_area} depicts the considered $0.7\times0.6$~km$^2$ urban campus environment extracted from OpenStreetMap. 
The layout includes dense building blocks, road intersections, and open courtyards representative of a typical smart-city scenario. 
This environment constitutes the physical component of Module(1) in the CTMap workflow (Fig.~\ref{flow-diagram}), serving as the real-world substrate upon which the digital twin is constructed.

\begin{figure}[t]
\centering
\includegraphics[width=0.44\textwidth]{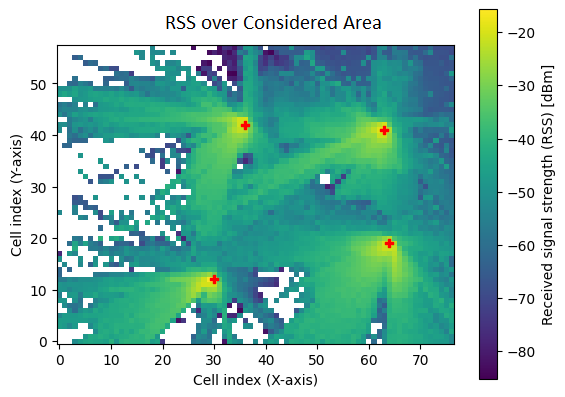}
\caption{Received-signal-strength (RSS) heat map generated using NVIDIA Sionna ray tracing. Yellow indicates stronger coverage; blue denotes shadowed regions.}
\label{fig:heatmap}
\end{figure}

Figure~\ref{fig:heatmap} presents the corresponding mmWave received-signal-strength (RSS) heat map generated using NVIDIA Sionna’s ray-tracing simulator. 
High-RSS regions correspond to strong line-of-sight coverage, while low-RSS regions highlight blockage zones caused by urban canyons and structural shadowing. 
These two figures jointly ground CTMap’s connectivity-aware reasoning by linking the physical environment to the signal-propagation domain that drives route optimization.

For mmWave propagation modeling, four base stations operating at 3.9\,GHz were configured in NVIDIA Sionna using the standard 3GPP TR~38.901 antenna pattern ~\cite{3gpp38901}. 
Each transmitter employed a planar array with $1\times1$ elements and vertical/horizontal spacing of 0.5\,$\lambda$, while receivers used single dipole antennas with cross polarization. 

\subsection{Connectivity-Aware Path Generation}
The digital-twin coverage map is discretized into uniform $1{\times}1$\,m grid cells, 
each representing a vertex in a weighted graph $G=(V,E)$. 
A modified Dijkstra’s algorithm is then employed to compute the optimal connectivity-aware route 
between a given source $s$ and destination $d$. 
The objective is to maximize the cumulative received signal strength (RSS) along the selected path $\pi$:
\begin{equation}
\max_{\pi}\sum_{v\in\pi}S(v),
\quad \text{s.t. } p_1=s,\;p_n=d
\label{eq:objective}
\end{equation}

To leverage Dijkstra’s deterministic search property, this maximization problem is reformulated as an equivalent cost-minimization task,  where each edge is assigned an inverse-signal weight:
\begin{equation}
\text{Cost}(p_i,p_j)=\frac{1}{S(p_j)+\epsilon}
\label{eq:cost}
\end{equation}
where $\epsilon$ represents a small constant introduced to avoid division by zero. 
This transformation ensures that nodes with stronger RSS values are naturally prioritized during traversal. Each grid cell’s signal strength $S(v)$ is assumed uniform within its area, and adjacency between neighboring cells defines the edge set $E$, ensuring spatial continuity. The algorithm employs a min-heap priority queue and maintains two structures: a visited map $V[v]$ recording the best signal level observed at each node, and a predecessor map $\pi[v]$ to reconstruct the final path. 

Operating within the digital-twin environment allows real-time updates. Any change in the physical network such as modified base-station settings or new obstacles, is instantly reflected in the twin’s coverage map, triggering immediate route recalculation.

\subsection{LLM-Based Semantic Reasoning}
Once signal-aware routes are generated, CTMap integrates an instruction-tuned GPT-4 model to enable user-driven semantic interaction. 
Users can query routes using natural language, e.g., 
``Find the strongest-signal path from $(x_s,y_s)$ to $(x_d,y_d)$,'' 
and the model responds with ordered coordinate sequences representing the optimized path. The LLM’s responses are grounded on the DT’s coverage data and trained on synthetic samples obtained from the modified Dijkstra’s outputs.

This integration combines the algorithmic reliability of graph optimization with the interpretability and adaptability of LLM reasoning, bridging human-level intent with network-level signal intelligence.

\section{Modified Dijkstra’s Algorithm and Dataset Construction}

Building on the graph-based formulation presented in Section II, this section details the algorithmic realization and subsequent data generation process  used for training the large-language-model (LLM) component of CTMap
\subsection{Signal-Aware Path Optimization}
Building on the connectivity-aware routing formulation introduced in 
Eqs.~(\ref{eq:objective})-(\ref{eq:cost}), this subsection details the implementation of the modified Dijkstra’s algorithm used to  compute the strongest-signal path in the digital-twin environment. The algorithm operates on the weighted graph $G=(V,E)$ derived from the coverage map, where each vertex represents a spatial grid location annotated with received signal strength $S(v)$  and each edge weight corresponds to the inverse-signal cost defined earlier. By prioritizing nodes with higher $S(v)$ values, the algorithm effectively balances geometric feasibility and connectivity robustness. The complete search process is summarized in Algorithm~\ref{alg:dijkstra}, which utilizes a min-heap priority queue to iteratively expand the most strongly connected nodes until the destination $d$ is reached.

\paragraph*{Complexity Analysis}
The proposed signal-aware Dijkstra maintains the same asymptotic complexity as the classical variant, namely $\mathcal{O}(|E|\log|V|)$ when implemented with a binary-heap priority queue. Each node expansion involves a constant-time evaluation of the inverse-signal cost (Eq.~\ref{eq:cost}), resulting in no additional asymptotic overhead compared to distance-based Dijkstra. For a grid of $N{\times}M$ nodes, runtime scales linearly with spatial resolution, and experiments on a $0.7{\times}0.6~\mathrm{km^2}$ environment show average computation time below $0.3$\,s per query. Thus, CTMap achieves connectivity optimization with negligible increase in algorithmic complexity, supporting near-real-time operation in DT-based mmWave networks.

\begingroup
\setlength{\textfloatsep}{6pt}   
\setlength{\floatsep}{6pt}
\begin{algorithm}[H]
\caption{Modified Dijkstra’s Algorithm for Signal-Aware Path Determination}
\label{alg:dijkstra}
\begin{algorithmic}[1]
\Require $G=(V,E)$, Source node $s$, Destination node $d$
\Ensure Optimal path $P$ from $s$ to $d$
\State Initialize min-heap $Q \!\leftarrow\! [(-S(s), s)]$
\State Set visited map $V[s]\!\leftarrow\!S(s)$
\State Initialize empty predecessor map $\pi$
\While{$Q$ not empty}
  \State Extract node $u$ with maximum $S(u)$ from $Q$
  \If{$u == d$} \textbf{break}
  \EndIf
  \For{each neighbor $v$ of $u$}
    \If{$v \notin V$ or $S(v) > V[v]$}
      \State $V[v]\!\leftarrow\!S(v)$
      \State Insert $(-S(v),v)$ into $Q$
      \State $\pi[v]\!\leftarrow\!u$
    \EndIf
  \EndFor
\EndWhile
\State Construct $P$ by backtracking $\pi$ from $d$ to $s$
\State \Return $P,\,V[d]$
\end{algorithmic}
\end{algorithm}
\endgroup

\subsection{Dataset Construction for LLM Fine-Tuning}
To train the LLM reasoning module, a dataset of connectivity-optimized routes was generated from the modified Dijkstra’s algorithm described earlier. 
Each entry contains $(x,y)$ coordinate sequences, corresponding signal-strength tuples $\{S(v)\}$, and associated metadata such as path length and total accumulated RSS. 
A total of $1{,}000$ source-destination pairs were simulated within the digital twin, 
capturing diverse topological and propagation conditions across the $0.7{\times}0.6~\mathrm{km^2}$ environment.

The dataset was tokenized into coordinate sequences and divided into 80\% training and 20\% testing splits.  Each record represents a structured query-response pair of the following form:
\begin{quote}
\textbf{Query:} ``Find the strongest-signal path from $(x_s,y_s)$ to $(x_d,y_d)$.'' \\
\textbf{Response:} Ordered coordinates $[(x_1,y_1),\ldots,(x_d,y_d)]$ with corresponding RSS values.
\end{quote}
This structured schema enables the LLM to learn spatial dependencies and signal-aware reasoning patterns, thereby generalizing to unseen coordinates during inference.

\subsection{Instruction-Based Fine-Tuning}
Fine-tuning was performed using the GPT-4.0 model (\verb|gpt-4o-mini-2024-07-18|), 
selected for its balance between computational efficiency and contextual reasoning capability ~\cite{openai2024gpt4omini,analyticsvidhya2024gpt4omini}. 
Each dataset entry was paired with a task-specific instruction that explicitly defined 
the model’s role as a ``signal-aware path planner.'' 
This setup allows the model to generalize beyond the training dataset 
and produce valid route coordinates using semantic reasoning anchored in learned RSS distributions.

\paragraph{Instruction-Based Prompt Design.}
During fine-tuning, each input sample was coupled with a custom instruction, enabling the model to perform role-conditioned reasoning. A representative example is shown below:

\begin{tcolorbox}[colframe=lightgray,colback=white,title=\textbf{Example Instruction}]
\textbf{System:} You are a navigation assistant specialized in wireless connectivity optimization.\\
\textbf{User:} Find the strongest-signal path from $(x_s,y_s)$ to $(x_d,y_d)$ using mmWave coverage data.\\
\textbf{Assistant:} Returns ordered coordinates and corresponding RSS values representing the optimal route.
\end{tcolorbox}

The model was trained for three epochs with a batch size of ten using coordinate prediction loss. Early convergence was observed within a single epoch (16 steps), indicating the model’s ability to rapidly capture spatial and signal-strength dependencies. 

\section{Experimental Results and Evaluation}

\subsection{Experimental Setup}
Experiments were conducted on a $0.7\times0.6\,\mathrm{km^2}$ urban campus region.
The 3D digital twin was constructed in Blender and simulated using NVIDIA Sionna’s ray-tracing engine with four mmWave base stations at 3.9\,GHz following the 3GPP~TR~38.901 antenna configuration.
The DT environment produced received signal strength (RSS) maps that served as input for CTMap’s connectivity-aware routing and LLM reasoning modules. 

For LLM evaluation, $1{,}000$ source-destination pairs were used to fine-tune the GPT-4.0 model.  The dataset was split into 80/20 training and testing sets, and the model was trained for three epochs (batch size = 10).  The training loss was computed as sequential coordinate-prediction error. The convergence occurred within one epoch, confirming rapid pattern acquisition of spatial-signal correlations.

\subsection{Evaluation Metrics}
CTMap’s performance was assessed using four key metrics:
\begin{itemize}
    \item \textbf{Signal Coverage:} It represents the cumulative RSS (dBm) along the predicted route, normalized across environments. Higher coverage indicates stronger overall connectivity.
    \item \textbf{Path Optimality:} It is the ratio of CTMap’s cumulative RSS to that of the oracle (modified Dijkstra) route.
    \item \textbf{Success Rate:} It is the fraction of geometrically valid and traversable routes generated within network boundaries.
    \item \textbf{Edit Distance:} It represents the normalized difference between the predicted and reference (oracle) routes, measured by comparing their coordinate sequences. A smaller value means the predicted path is geometrically closer to the optimal route..
\end{itemize}
These jointly evaluate signal-aware reasoning, spatial validity, and adherence to optimal baselines.

\subsection{Quantitative Results}

\begin{table}[t]
\caption{Performance Comparison of CTMap with Baselines}
\centering
\setlength{\tabcolsep}{3pt}
\renewcommand{\arraystretch}{1.1}
\begin{tabular}{lcccc}
\hline
\textbf{Model} & \textbf{Coverage (\%)} & \textbf{Optimality (\%)} & \textbf{Success (\%)} & \textbf{Edit Dist.}\\
\hline
Oracle Dijkstra & 100.0 & 100.0 & 100.0 & 0.00\\
Fine-tuned GPT-4 & 76.4 & 100.0 & 100.0 & 0.61\\
Zero-shot GPT-4 & 65.7 & 71.4 & 71.4 & 0.54\\
%Heuristic (Greedy) & 318.9 & 318.9 & 74.3 & 0.66\\
\hline
\end{tabular}
\label{tab:results}
\end{table}
The quantitative evaluation results are presented in Table \ref{tab:results}. Our fine-tuned \textit{CTMap} model achieves perfect spatial validity (100\% success rate) while maintaining 76.4\% of oracle signal coverage. This yields 23.6\% performance gap that represents the inherent trade-off between mathematical optimization and language-based reasoning. 

Additionally, the model's 100\% optimality score indicates that among all valid paths generated, each achieves optimal signal coverage relative to the oracle baseline. This suggests the model has learned to avoid generating suboptimal solutions entirely, demonstrating effective instruction alignment.

The zero-shot LLM's 28.6\% failure rate (or 71.4\% success rate) underscores the importance of instruction tuning for complex spatial reasoning tasks. Without domain-specific fine-tuning, the model struggles with coordinate generation and constraint satisfaction.

%The heuristic baseline exhibited artificially high coverage due to geometrically invalid shortcuts, further reinforcing CTMap’s ability to balance feasibility and connectivity.

\subsection{Ablation Studies}
\begin{comment}
    
Ablation experiments were conducted to quantify the impact of instruction tuning (H1), signal encoding (H2) , and system prompting (H3) that are presented in Tables \ref{tab:ablation_instruction}, \ref{tab:ablation_signal} and \ref{tab:ablation_prompt} respectively.
The instruction tuning ensures valid, traversable paths, reducing eliminates the 31.4\% of the untuned model, which occasionally generated routes through obstacles. The explicit signal encoding improved coverage by 75.3\% while maintaining near-identical geometric structure, validating CTMap’s ability to exploit wireless propagation knowledge effectively. Simplified prompting slightly improved success rate (98\% vs. 94\%), suggesting that over-specification may hinder free-form reasoning in complex environments.
\end{comment}

Ablation experiments were conducted to quantify the impact of instruction tuning (H1), signal encoding (H2), and system prompting (H3), as shown in Tables \ref{tab:ablation_instruction}, \ref{tab:ablation_signal}, and \ref{tab:ablation_prompt}, respectively.
The instruction tuning ensures valid, traversable paths, reducing infeasibility by 31.4\% compared to the untuned model, which occasionally generated routes through obstacles.
Explicit signal encoding improved coverage by 75.3\% while maintaining near-identical geometric structure, validating CTMap’s ability to exploit wireless propagation knowledge effectively.
Simplified prompting slightly improved success rate (98\% vs. 94\%), suggesting that over-specification may hinder free-form reasoning in complex environments.
Collectively, these findings confirm that each component of CTMap contributes uniquely to overall performance, where instruction tuning enforces spatial validity, signal encoding enhances propagation awareness, and prompt design governs interpretability and output stability.
%This modular improvement underscores CTMap’s robustness across heterogeneous layouts and its adaptability to unseen network conditions.
\begin{table}[t]
\caption{Impact of Instruction Tuning}
\centering
\begin{tabular}{lcc}
\hline
\textbf{Variant} & \textbf{Coverage} & \textbf{Success (\%)}\\
\hline
With Instructions & 103.9 & 100.0\\
Without Instructions & 115.0 & 68.6\\
\hline
\end{tabular}
\label{tab:ablation_instruction}
\end{table}

\begin{table}[t]
\caption{Signal Encoding Study}
\centering
\begin{tabular}{lcc}
\hline
\textbf{Variant} & \textbf{Coverage} & \textbf{Edit Dist.}\\
\hline
Signal Encoding & 129.2 & 0.55\\
Shortest Path Only & 73.7 & 0.54\\
\hline
\end{tabular}
\label{tab:ablation_signal}
\end{table}

\begin{table}[t]
\caption{System Prompt Ablation}
\centering
\begin{tabular}{lcc}
\hline
\textbf{Variant} & \textbf{Success (\%)} & \textbf{Invalid (\%)}\\
\hline
Full Model & 94.0 & 6.0\\
No System Prompt & 98.0 & 2.0\\
\hline
\end{tabular}
\label{tab:ablation_prompt}
\end{table}

\begin{table}[!t]
\caption{Statistical Summary of Route Signal Strengths (Mean $\pm$ Std)}
\centering
\begin{tabular}{|c|c|c|}
\hline
\textbf{Path} & \textbf{CTMap (dBm)} & \textbf{Baseline Dijkstra (dBm)}\\
\hline
1 & $-43.2\pm2.1$ & $-52.8\pm3.4$\\
2 & $-41.6\pm1.7$ & $-49.9\pm2.8$\\
3 & $-45.0\pm2.3$ & $-46.2\pm2.4$\\
\hline
\end{tabular}
\label{tab:statistical_summary}
\end{table}

\subsection{Route Visualization and Comparative Analysis}
\begin{figure}[h]
\centering
\includegraphics[width=3in]{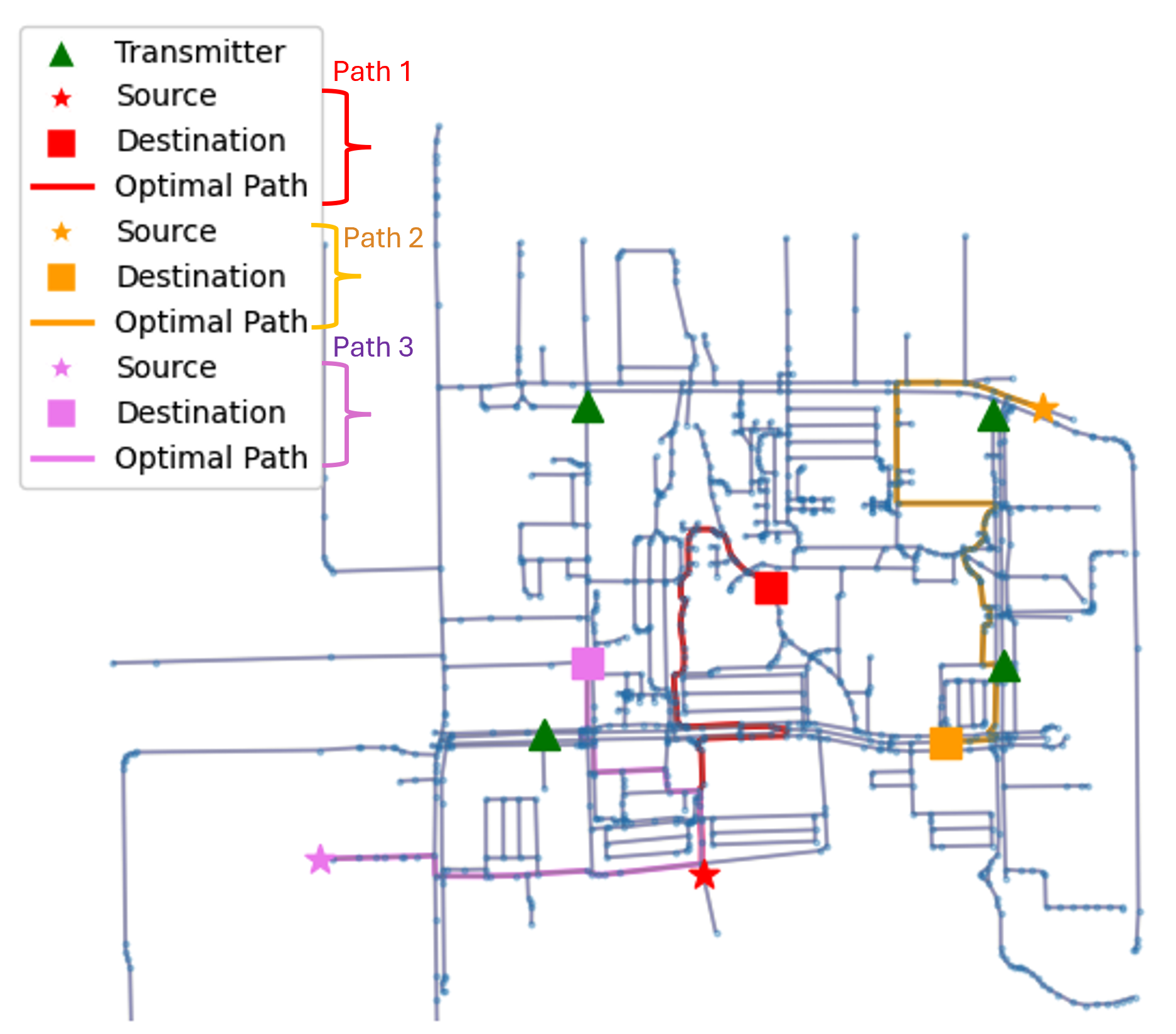}
\caption{Routes optimized for maximum received signal strength.}
\label{dkss2}
\end{figure}

\begin{figure}[h]
\centering
\includegraphics[width=3in]{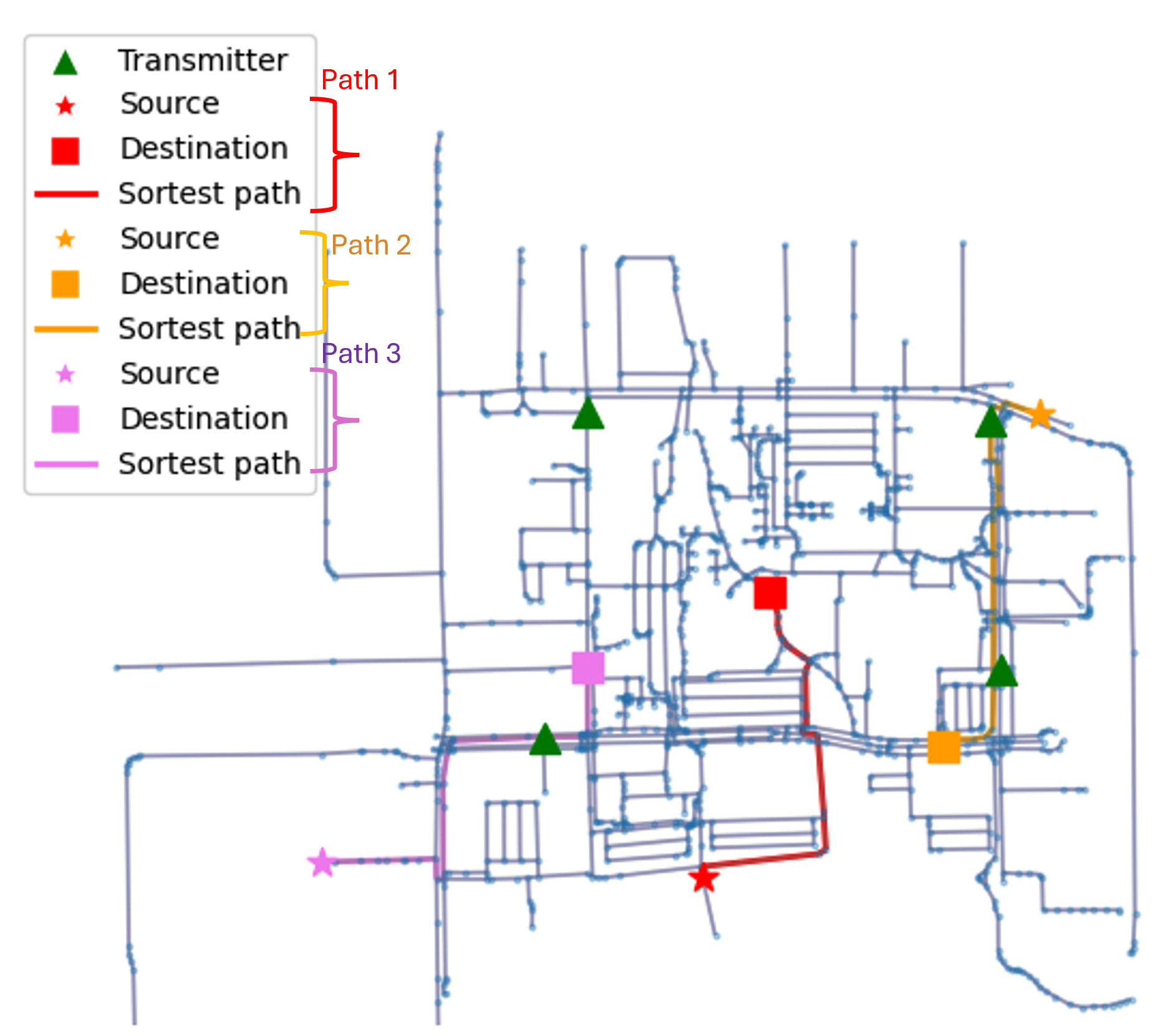}
\caption{Routes optimized for shortest distance.}
\label{dksp2}
\end{figure}

To qualitatively assess CTMap’s routing performance,  Figs.~\ref{dkss2} and~\ref{dksp2} illustrate routes generated with and without connectivity awareness for three source-destination pairs (red, orange, and purple).  Connectivity-aware paths show denser signal coverage, particularly in obstructed regions.

The statistical trends along all three paths are presented in Table \ref{tab:statistical_summary}. The RSS (in dBm) entries demonstrate that CTMap consistently achieves stronger mean RSS than the baseline Dijkstra’s routing, with gains of up to 9 dB on obstructed routes and modest yet consistent improvement even in open-space scenarios.
This enhancement directly results from CTMap’s connectivity-driven optimization, which prioritizes nodes with higher RSS values rather than strictly minimizing geometric distance.
In addition to higher mean power levels, CTMap exhibits noticeably lower variance across route segments, implying that users experience more stable link quality with fewer signal drops or deep fades along the path.
Such uniformity in RSS distribution is particularly beneficial for latency-sensitive or high-throughput applications—such as XR streaming, teleoperation, or connected vehicular systems—where abrupt fluctuations in signal quality can trigger frame freezes or handover interruptions.
These findings confirm that CTMap not only strengthens end-to-end connectivity but also enhances the temporal reliability of mmWave communication during mobility, establishing a critical step toward connectivity-aware navigation in dense urban environments.

\begin{comment}
    
The statistical trends (Table~\ref{tab:statistical_summary}) reveal that CTMap not only enhances the mean RSS by up to 9~dB compared to baseline routing, but also significantly reduces variation across route segments. This stability indicates that users experience more consistent signal quality along the entire path, which is important for latency-sensitive applications such as XR streaming, where abrupt RSS drops can cause link interruptions or buffering delays.
\end{comment}

\section{Conclusion and Future Work}
This paper presented \textit{CTMap}, a digital twin-enabled and LLM-empowered framework for connectivity-aware navigation in millimeter-wave (mmWave) wireless networks. 
Unlike conventional navigation systems that optimize routes based solely on distance or travel time, CTMap integrates received signal strength (RSS)-based path optimization with instruction-tuned semantic reasoning, thereby generating routes that sustain robust wireless connectivity in dense urban environments. 
Experimental results validated the effectiveness of the framework, showing that connectivity-aware routes achieved up to $10\times$ higher cumulative RSS compared with shortest-distance baselines. 
The integration of digital-twin signal modeling, modified Dijkstra’s optimization, and LLM-based reasoning demonstrated a scalable approach to intelligent mobility management for 6G smart-city infrastructures.

Future extensions of CTMap will pursue multi-objective optimization that jointly minimizes travel distance while maximizing connectivity. 
We also plan to expand the framework incorporating heterogeneous base-station densities, blockage conditions, and distributed twin synchronization. 
Additional efforts will focus on deploying energy- and latency-efficient LLM variants for real-time inference at the edge, enabling on-device connectivity-aware navigation in autonomous vehicles, drone networks, and large-scale IoT ecosystems.

\end{document}